\documentclass[12pt]{article}
\usepackage[pdftex]{graphicx}
\usepackage[pdftex]{color}
\usepackage{times}
\usepackage{hyperref}

\newcommand{\beq}{\begin{equation}}
\newcommand{\eeq}{\end{equation}}

\def\E{\ensuremath {\cal E}}

\title{TRI-BN-25-6: Cyclotron Spiral Inflector}
\author{Rick Baartman, TRIUMF}
\date{March, 2025}
\begin{document}
\maketitle
\raggedright
\begin{abstract}
The most common device for injecting into a compact cyclotron is an electrostatic device known as the spiral inflector. There is a known analytic solution for the trajectory, and therefore the electrode shape, in the idealized case of a uniform magnetic field. This is re-derived in this note and the conditions for perfect centring are found. It is moreover found that the `M\"{u}ller', or hyperboloid inflector is actually a special case of the spiral inflector.\end{abstract}

\section{Introduction}
The function of the inflector is to take the beam from the cyclotron's axis to the median plane while its velocity is initially aligned on the axis and finally also in the median plane. Positionally, we would like to place the beam on the closed orbit corresponding to its energy. This orbit has radius $\rho$ given by $qB\rho=mv$, $B$ is the central field of the cyclotron.

Invented by Belmont and Pabot\cite{belmont1966study}, the spiral inflector is most simply understood as a $90^\circ$ cylindrical electrostatic bend that has been modified to keep the electric force perpendicular to the beam while the beam twists and spirals due to the cyclotron's field. The electric field strength (${\cal E_{\rm inf}}=\frac{V_{\rm o}-V_{\rm i}}{x_{\rm sep}}$, the voltage difference between the plates divided by their separation) is the parameter that directly determines the inflector height $A=\frac{2V_0}{\cal E_{\rm inf}}$ ($V_0$ is the beam's energy per charge, i.e.\ the voltage of extraction from the ion source). 

\section{No tilt}\subsection{Electrostatic $90^\circ$ bend}
Let us start by ignoring the magnetic field. Let $z$ be the axis along which the particle's initial velocity is aligned, and let $x$ be the direction of the electric field at the entrance. We choose the cyclotron centre as origin, the electric field is everywhere orthogonal to the velocity, so the trajectory is on the equipotential surface $(x-A)^2+(z-A)^2=A^2$. The electric field is in the $x$-$z$ plane and radial component (`radial' in $xz$, not radial in the median plane $xy$) is 
\beq 
q{\E}_r=mv^2/A=2qV_0/A:=q{\E}_0.
\eeq  
$V_0$ is the beam energy per unit charge: the bias voltage of the source terminal in most cases. In the Cartesian frame, the electric field components are 
\beq 
{\E}_x={\E}_0\cos(s/A),{\E}_z={\E}_0\sin(s/A),
\eeq 
where $s=v_0t$ is the distance along the trajectory, running from zero to $\frac{\pi}{2}A$. The equations of motion are simply 
\beq 
x''=\frac{1}{A}\cos b,y''=0,z''=\frac{1}{A}\sin b.
\eeq 
Primes are derivatives with respect to $s$, and to clean the notation, I introduce $b:=s/A$. Solutions for initial conditions $x=y=x'=y'=0,z'=-1$ are: 
\beq 
x'=\sin b,y'=0,z'=-\cos b,
\eeq 
\beq x=A-A\cos b,y=0,z=A-A\sin b.
\eeq This is also consistent with the equipotential surface $(x-A)^2+(z-A)^2=A^2$. 

\subsection{Add the magnetic field}With the magnetic field added, there is a magnetic force $\vec{v}\times\vec{B}$, which adds to $x''$ a term $y'/\rho$, and to $y''$ a term $-x'/\rho$. The electrodes must be modified to maintain the electric force orthogonal to $\vec{v}$. The result is a spiral shape for the trajectory and the electrodes. In this configuration, it remains the function of the electric field to rotate the velocity vector from vertical to horizontal; the magnetic field provides the only radial ($xy$) force. The electric field that does this is 
\beq 
{\E}_x={\E}_0\cos kb\cos b,{\E}_y=-{\E}_0\sin kb\cos b,
\eeq 
while ${\E}_z$ is unmodified. We have introduced $k:=A/\rho$. In this way, the field orthogonal to $z$-axis remains ${\E}_0\cos(s/A)$, but is divided between $x$ and $y$ according to the needs of the rotation caused by the magnetic field. The magnitude of the electric field is unchanged from the cylindrical bend, as ${\E}_0^2={\E}_x^2+{\E}_y^2+{\E}_z^2$.

The equations of motion are now:
\begin{eqnarray}
x''&=&\frac{1}{A}\cos kb\cos b+\frac{y'}{\rho},\nonumber\\
y''&=&\frac{-1}{A}\sin kb\cos b-\frac{x'}{\rho},\label{eq.force}\\
z''&=&\frac{1}{A}\sin b.\nonumber
\end{eqnarray}
These are the solutions from a single integration:
\begin{eqnarray}
x'&=&\ \cos kb\sin b,\nonumber\\
y'&=&-\sin kb\sin b,\label{eq.mimic}\\
z'&=&-\cos b.\nonumber
\end{eqnarray}
The reader is invited to verify both that the velocity $\vec{v}$ is invariant, and that $\vec{v}\cdot\vec{{\E}}=0$. 

One more integration yields the equations originally derived (but in a slightly different form)\cite{belmont1966study}:
\begin{eqnarray}
x &=&\frac{A}{k^2-1}\left[k\sin kb\sin b+\cos kb\cos b\,-1\right],\nonumber \\
y &=&\frac{A}{k^2-1}\left[k\cos kb\sin b-\sin kb\cos b\right],\label{eq.eom}\\
z &=&A-A\sin b.                                     \nonumber
\end{eqnarray}
(It may look as though the $k=1$ case is singular but this is not so; it is a removable singularity.)

\begin{figure}[htbp]
\begin{center}
\includegraphics[width=\textwidth]{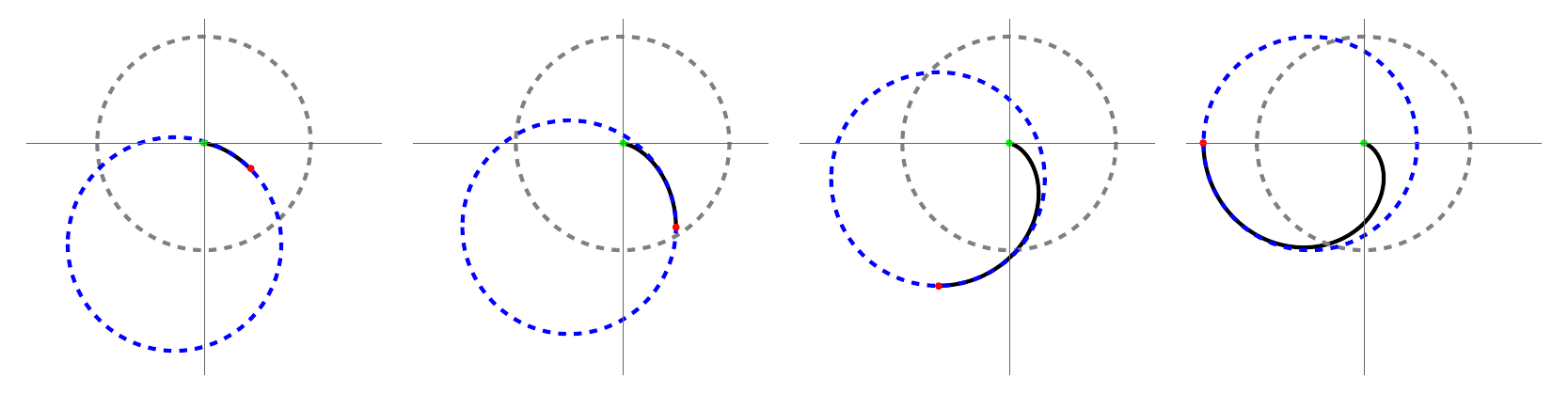}
\caption{Overhead views of four untilted ($k'=0$) inflectors assuming beam enters the inflector on the cyclotron's vertical axis (green dot). Left to right: $A/\rho=0.5,1,2,3$. In all cases, the gray orbit is the intended one, and the blue orbit is the orbit placed by the inflector. In each case, the initial deflection is in the $+x$ direction, and that this rotates by $\frac{A}{4\rho}$ turn entry to exit.}
\label{fig:untiltOH}
\end{center}
\end{figure}

As described so far, this inflector does not have sufficient flexibility to place the beam on the correctly centred closed orbit. Only the height $A$ is variable. Figure \ref{fig:untiltOH} shows a few examples as seen from above, i.e., as projected onto the median plane. Note that for the $A/\rho=1$ case, the beam arrives on the orbit, but travelling in the wrong direction. As $A/\rho\rightarrow\infty$, the beam becomes better centred. One can show that this inflector places the beam correctly only in the limit $A\gg\rho$. In this limit the beam takes many turns of the spiral (specifically, $\frac{A}{4\rho}$ turns) to reach the median plane. Such a tall inflector is impractical for many reasons, not least of which is that there is no vertical focusing for this long path. In any case, the idealized case of a uniform axial magnetic field independent of $z$ as considered here becomes less tenable as the height becomes much larger than $\rho$. In general $A$ taller than about $2\rho$ is unacceptable as it makes the inflector too large. At the small end of the scale, the beam is placed very near the centre of the cyclotron and not on the orbit, and in those cases the electric field is much higher than need be.

\section{With tilt}
To achieve adjustability for the placement, another parameter is needed and this is provided by tilting the electrodes. The intent is to add more electric field in the $x$-$y$ direction without altering the $z$ component. If the electric force in the $x$-$y$-plane is made to mimic the magnetic force then the equations of motion will remain unchanged. In reference to eqns.\,\ref{eq.force},\ref{eq.mimic}, we see that the added electric field must have the form $({\E}_x,{\E}_y)\propto(\cos kb\sin b,-\sin kb\sin b)$. This is achieved by tilting the electrodes and narrowing the gap. The tilt $\theta_{\rm t}$ is the angle between the electric field and the vector formed from the cross product between the beam's  velocity and the orthogonal of this vector that lies parallel to the median plane. This tilt is applied gradually according to \beq \tan\theta_{\rm t}=k'\sin b.\eeq In this way, the vertical component of the electric field remains as for the untilted case, but an additional radial component is gradually applied; while still orthogonal to the beam, the electric force can augment or diminish the magnetic force, thus tightening or loosening the orbit's curvature.

\begin{figure}[htbp]
\begin{center}
\includegraphics[width=\textwidth]{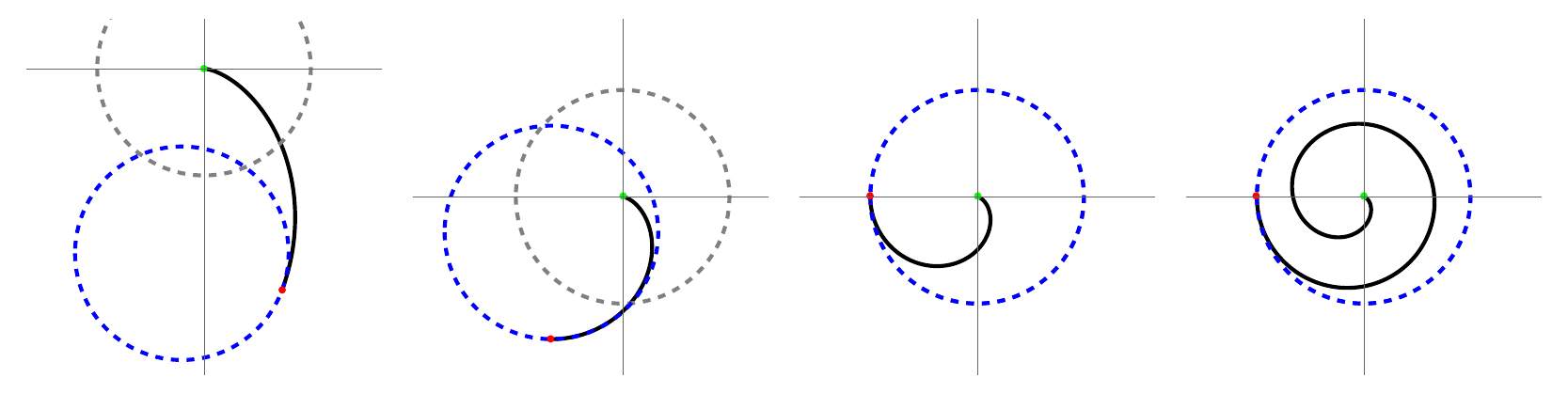}
\caption{Overhead views of four inflectors assuming beam enters the inflector on the cyclotron's vertical axis (green dot). Left to right: the hyperboloid ($A/\rho=\sqrt{6}$, $k'=-\sqrt{6}/2$), the zero tilt ($k'=0$) with $A/\rho=2$, the ideally centred ($A/\rho=2$, $k'=1$) case, and lastly, the ideally centred tall case where $A/\rho=6$, $k'=1$. In all cases, the gray orbit is the intended one, and the blue orbit is the orbit placed by the inflector.}
\label{fig:spiral_OH}
\end{center}
\end{figure}

The $x$ and $y$ equations of motion ($z$ is unchanged) become:
\begin{eqnarray}
x''&=&\frac{1}{A}\cos kb\cos b-k'\sin kb\sin b+\frac{y'}{\rho}=\nonumber\\
   & &=\frac{1}{A}\left[\cos kb\cos b-k\sin kb\sin b\right],\\
y''&=&\frac{-1}{A}\sin kb\cos b-k'\cos kb\sin b-\frac{x'}{\rho}=\nonumber\\
   & &=\frac{-1}{A}\left[\sin kb\cos b+k\cos kb\sin b\right],\label{eq.force2}\nonumber
\end{eqnarray}
and the parameter $k$ has now been redefined as
\begin{equation}
  k=\frac{A}{\rho}+k'.
\end{equation}
With this redefinition of $k$, the solutions are identical to eqns.\,\ref{eq.mimic},\ref{eq.eom}. (Note that this definition of $k$ is different by a factor of 2 from the definition of $K$ used by Belmont and Pabot\cite{belmont1966study} and by Root\cite{Root_1972}, and also Goswami et al.\cite{goswami2019design}. Also, these equations agree with the derivation of Goswami et al.\ but for a different sign convention of $B$.) This parameter turns out to be the number of quarter turns that the beam's velocity vector makes through the inflector, as projected onto the median plane (see Fig.\,\ref{fig:spiral_OH}).

\begin{figure}[htbp]
\begin{center}
\includegraphics[width=.85\textwidth]{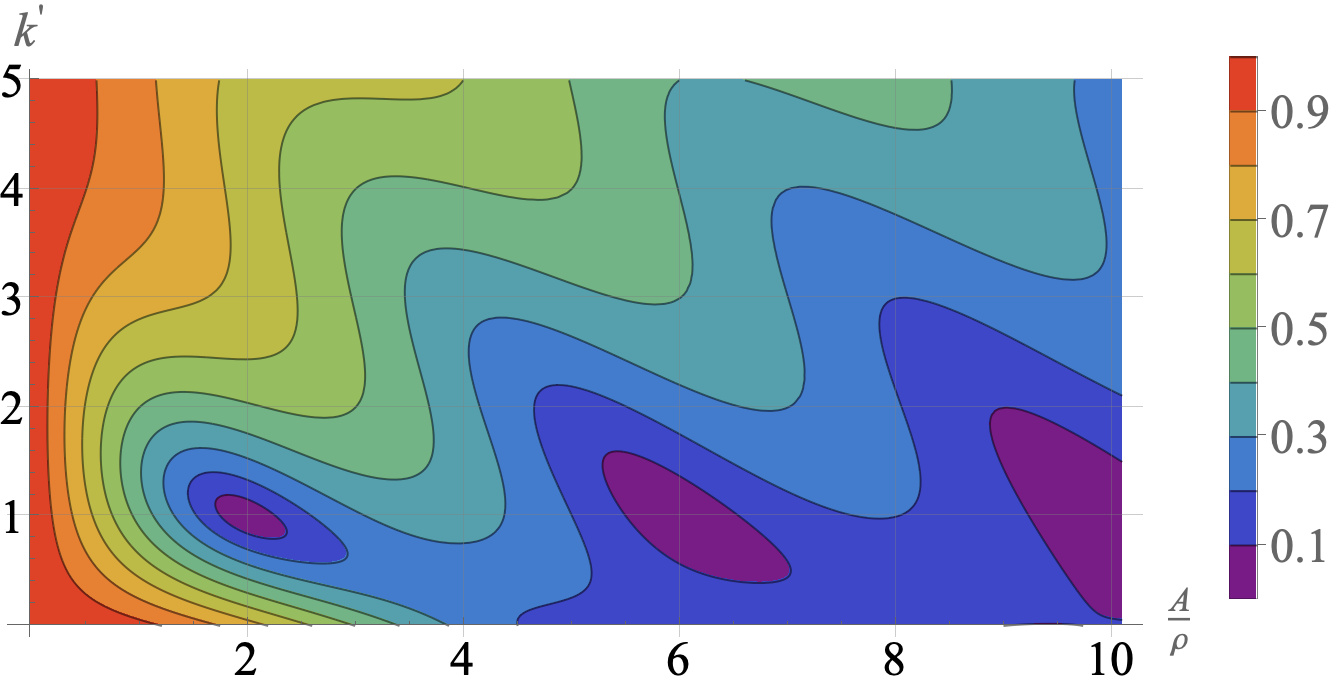}
\caption{The centring error of the orbit placed by the spiral inflector, in units of the orbit radius, as a function of electric radius $A$ and tilt parameter $k'$.}
\label{fig:cvsAkp}
\end{center}
\end{figure}
The reader can show that the beam is placed precisely on the correct closed orbit when this inflector has the two parameters at $A/\rho=2$ and $k'=1$. This is shown in Fig.\,\ref{fig:spiral}\footnote{See Appendix for note on how to view in 3D using such stereoscopic pairs.}.
\begin{figure}[htbp]
\begin{center}
\includegraphics[width=.5\textwidth]{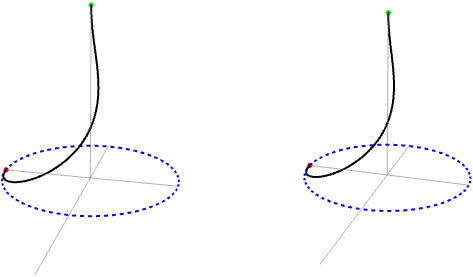}
\caption{The beam's trajectory for the spiral inflector for the case of perfect centring ($A/\rho=2$ and $k'=1$). This is a stereoscopic view. The initial horizontal deflection is to the right ($+x$) and finally is directly away from the viewer, making a total sweep of $k\pi/2=3\pi/2$ radians or $3/4$ turn. The blue dashed circle is the closed orbit, and found centred at the origin, which is the cyclotron centre.}
\label{fig:spiral}
\end{center}
\end{figure}
Interestingly, there are other perfectly centred solutions at $k'=\pm1$, and $\frac{A}{2\rho}$ is any odd integer ($A/\rho=2,6,10,...$, see Fig.\,\ref{fig:cvsAkp}). (Those with negative $k'$ can be ignored as they unnecessarily increase the inflector size laterally.) These have not been explored. Larger electric radius $A$ means that the electric fields are lower, but on the other hand, such solutions require a very tall inflector, over which the magnetic field is to be uniform. As stated above, making room for such an inflector is problematic, considering the flatness of the magnetic field needed on the median plane to keep the beam focused and isochronous.

The hyperboloid inflector\cite{muller1967novel} is just a special case of the tilted spiral inflector. Its parameters are: $A/\rho=\sqrt{6}$ and $k'=-\sqrt{6}/2$. From the figure, one can see the difficulty of implementing this inflector: the tilt parameter has the wrong sign, working in opposition to the magnetic force.  \begin{figure}[htbp]
\begin{center}
\includegraphics[width=.5\textwidth]{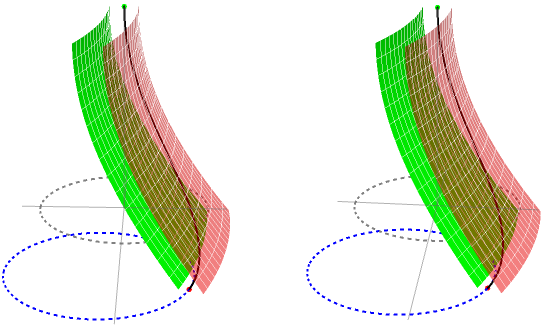}
\caption{The two surfaces of the electrodes of the hyperboloid inflector, including the beam's trajectory. The inner electrode surface is green, and the outer is red but shown as semi-transparent to make visible the trajectory. This is a stereoscopic view. The initial horizontal deflection is to the right ($+x$) and finally comes toward the viewer with a slight leftward slant, making a total sweep of $k\pi/2=\sqrt{6}\pi/4\approx110^\circ$. The dashed circles are in the $x$-$y$-plane; the gray one is the centred orbit and the blue one is the actual orbit if uncorrected.}
\label{fig:hyperboloid}
\end{center}
\end{figure}

\begin{figure}[htbp]
\begin{center}
\includegraphics[width=.5\textwidth]{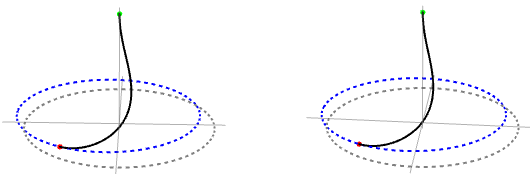}
\caption{The beam's trajectory for the spiral inflector for the TRIUMF case $A/\rho=1.2$ and $k'=1.09$. This is a stereoscopic view. The blue dashed circle is the closed orbit as placed by the inflector. To correct the orbit centring, an electrostatic deflector is placed just downstream of the inflector, where the blue orbit crosses the gray centred orbit. (Note that actually the TRIUMF beam rotates anti-clockwise $(A/\rho,k')=(-1.2,-1.09)$, but shown here as clockwise to be consistent with the other plots.)}
\label{fig:TRIUMFspiral}
\end{center}
\end{figure}

Perfect orbit placement can however be achieved even without conforming to the centring conditions ($A/\rho=2$ and $k'=1$); one can compensate by providing additional horizontal steering. For example in the TRIUMF case, $A/\rho=2$ would require a height of $A=51$\,cm (since $\rho=25.4$\,cm) which is too large and would interfere with the magnet steel at the cyclotron centre. Choosing $A/\rho=1.2$ and $k'=1.09$ allowed the inflector to be only $30$\,cm in height, and the centring error could be corrected with a separate horizontal deflector 6.5\,cm downstream of the inflector exit. This is shown in Fig.\,\ref{fig:TRIUMFspiral}; the deflector is located where the placed beam's orbit and the centred orbit cross.

The four cases -- hyperboloid, zero tilt, ideally centred, and ideally centred with large $A$ -- are shown in Fig.\,\ref{fig:spiral_OH}. This clarifies the function of the electrode tilt.

\section{Non-ideal cases}
By `ideal' is meant that the magnetic field is uniform over the region of the inflector and the first turn in the cyclotron. This is often not the case, as the magnetic field begins to decrease going backward up the axis farther than one half the magnet gap from the median plane. 

Secondly, the inflector described above, has a vertical electric field ${\E}_z=2V_0/A$ while also supplying an inward radial force that is $k'$ times the vertical one. For this to be true the electrode gap must narrow by a factor $1/\cos\theta_{\rm t}=\sqrt{1+k'^2\sin^2(s/A)}$ with distance $s$ along the inflector. The advantage of this configuration is that the orbit can be expressed analytically. In actual practice, the gap is left constant as that will minimize the electric field. The disadvantage is that this makes the inflector somewhat taller, and also the electrode shape can only be found by iterating between a Laplace solver and a particle tracker code. But this latter will be true in most cases anyway, because of the non-uniformity of the magnetic field.

Still, the idealized case allows a very good first step toward a cyclotron design.

\bibliographystyle{elsarticle-num}
\bibliography{/Users/baartman/AllDN/Baartman,/Users/baartman/AllDN/AllDN,/Users/baartman/AllDN/Others}

\section*{Appendix: 3D viewing using stereoscopic pairs}
The pairs of figures are to enable you, the reader to view a 3-dimensional image. This is done by having two views from slightly different angles as would occur when your two eyes actually see a real 3D object. Note that the figures here are in a `parallel-view', not a `cross-view', which would requiring crossing one's eyes. 

To view them, imagine the object is far in the distance so that you see 4 images (2 from each eye), and convince the eyes to superimpose the centre two. Once that is achieved, try to focus without re-converging the eyes. You may need to adjust the magnification on screen so that the two images are no farther apart than your PD (pupillary distance of your eyes).

An alternative and more foolproof method is to place a barricade such as a sheet of paper between the eyes and extending to the screen or page, between the images; the intent is that the left eye see only the left image and the right eye see only the right image.

\end{document}